# Structural Transitions and Energy Landscape for Cowpea Chlorotic Mottle Virus Capsid mechanics from nanomanipulation *in vitro* and *in silico*[*]

Olga Kononova,[†‡] Joost Snijder,[§] Melanie Brasch,[¶] Jeroen Cornelissen,[¶] Ruxandra I. Dima,[‖] Kenneth A. Marx,[†] Gijs J. L. Wuite,[§] Wouter H. Roos[§*] and Valeri Barsegov[†‡*]

[†]Department of Chemistry, University of Massachusetts, Lowell, MA 01854; [‡]Moscow Institute of Physics and Technology, Moscow Region, Russia 141700; [§]Natuur- en Sterrenkunde and LaserLab, Vrije Universiteit, 1081 HV Amsterdam, The Netherlands; [¶]Biomoleculaire Nanotechnology, Universiteit Twente, 7500 AE Enschede, The Netherlands; [‖]Department of Chemistry, University of Cincinnati, Cincinnati, OH 45221.

*Running Title*: Mechanical properties of CCMV capsid

[*]Corresponding authors:
e-mail: **Valeri_Barsegov@uml.edu**, tel: 978-934-3661, fax: 978-94-3013
e-mail: **wroos@few.vu.nl**, tel: +31 20 59 83974, fax: +31 20 59 87991

## ABSTRACT

Physical properties of capsids of plant and animal viruses are important factors in capsid self-assembly, survival of viruses in the extracellular environment, and their cell infectivity. Virus shells can have applications as nanocontainers and delivery vehicles in biotechnology and medicine. Combined AFM experiments and computational modeling on sub-second timescales of the indentation nanomechanics of Cowpea Chlorotic Mottle Virus (CCMV) capsid show that the capsid's physical properties are dynamic and local characteristics of the structure, which depend on the magnitude and geometry of mechanical input. Surprisingly, under large deformations the CCMV capsid transitions to the collapsed state without substantial local structural alterations. The enthalpy change in this deformation state $\Delta H = 11.5 - 12.8$ MJ/mol is mostly due to large-amplitude out-of-plane excitations, which contribute to the capsid bending, and the entropy change $T\Delta S = 5.1 - 5.8$ MJ/mol is mostly due to coherent in-plane rearrangements of protein chains, which result in the capsid stiffening. Dynamic coupling of these modes defines the extent of elasticity and reversibility of capsid mechanical deformation. This emerging picture illuminates how unique physico-chemical properties of protein nanoshells help define their structure and morphology, and determine their viruses' biological function.

## INTRODUCTION

Hierarchical supramolecular systems that spontaneously assemble, disassemble and self-repair play fundamental roles in biology. Prime examples are viral capsids, which exhibit pronounced features such as shape alteration, viscoelasticity, and materials fatigue. Understanding the microscopic structural origin of the physico-chemical properties of these biological assemblies and the mechanisms of their response to controlled mechanical inputs, remains a key research challenge. Beyond a fundamental understanding of these issues, there is a rapidly increasing interest in protein nanoshells for applications in biotechnology and nanomedicine. Recently, single-molecule techniques, such as Atomic Force Microscopy (AFM), have become available to explore physical properties of biological assemblies (1,2). This triggered tremendous research effort to characterize a variety of protein shells of plant and animal viruses, and bacteriophages. AFM based mechanical testing of viral capsids has now become the main tool to probe the physico-chemical and materials properties of viruses (3). For example, AFM deformation experiments yield information on the particle spring constant, reversibility of deformation, and forces required to distort capsid structures.

A variety of viruses have been tested by this method including the bacteriophages $\Phi 29$, $\lambda$ and *HK*97 (4-6), the human viruses Human Immunodeficiency Virus, Noro Virus, Hepatitis B Virus, Adeno Virus and Herpes Simplex Virus (7-13) and other eukaryotic cell infecting viruses like Minute Virus of Mice, Triatoma Virus and Cowpea Chlorotic Mottle Virus (14-16). These experiments reveal a truly amazing diversity of mechanical properties of viruses. Yet, due to their high complexity ($\sim 10^4$-$10^5$ amino acid residues), experimental results are difficult to interpret without input from theoretical modeling. For biotechnological applications, it is essential to have full control over structure-based physical properties of virus shells, but in most instances a detailed knowledge of these properties is lacking.

Viral capsids possess modular architectures, but strong capsomer intermolecular couplings modulate their properties. Consequently, the properties of the whole system (capsid) might not be given by the sum of the properties of its structural units (capsomers) (17-20). Under these circumstances, one cannot reconstruct the mechanical characteristics of the whole system using only information about the physical properties of its components. Biomolecular simulations have become indispensable for the theoretical exploration of the important dynamical properties and states of biological assemblies (21-23). Yet, the large temporal band width (ms-s) required limits the current theoretical capabilities. Theoretical studies employing triangulation of spherical surfaces and bead-spring models of stretching and bending have been used to probe the mechanical deformation and to test the mechanical limits of the shell (24,25). Yet, questions remain concerning structural details and dynamical aspects of these properties. How do discrete microscopic transitions give rise to the continuous mechanical response of a capsid at the macroscopic level? What are the structural rearrangements that govern the capsid's transition from the elastic to plastic regime of the mechanical deformation?

To address these questions, alternative detailed simulation approaches are needed. Here we feature a computational model, which is based upon the notion that the unique features associated with native topology, rather than atomic details, govern the physico-chemical properties of virus capsids. Our approach employs a topology based Self Organized Polymer (SOP) model (26,27), which provides an accurate description of the polypeptide chain (28-30), and high-performance computing accelerated on Graphics Processing Units (GPUs) (31,32). Here, we show that by combining AFM-based force measurements with accurate biomolecular simulations of forced indentation, we can obtain an in-depth understanding of the structural transitions and mechanisms of the mechanical deformation and the transition to the collapsed state in virus shells. On the theoretical side, the main challenge is to generate long (0.01 s − 0.1 s) dynamics of a virus particle using experimentally relevant force loads. Here, for the first time, we directly compare the results of experiments and simulations obtained under similar conditions of force application for the Cowpea Chlorotic Mottle Virus (CCMV) used as a model system (15,33-38).

CCMV is a member of the *Bromoviridae*, an important family of single stranded RNA plant viruses distributed worldwide that infect a range of hosts and are the cause of some major crop epidemics (39). The capsid of CCMV is an icosahedral protein shell (triangulation number $T = 3$) with an outer radius $R = 13.2$ nm and average shell thickness of 2.8 nm (3,40) consisting of 180 copies of a single 190 amino acid long protein. The shell comprises 60 trimer structural units and exhibits pentameric symmetry at the 12 vertices (pentamer capsomeres) and hexameric symmetry at the 20 faces (hexamer capsomeres) of the icosahedron (Fig. 1). The excellent agreement we demonstrated between experiments and simulations allowed us to describe the structural properties of the CCMV capsid, devoid of nucleic acids, to unprecedented detail, and to link the structural transitions in a protein shell with its mechanical properties. The insights into the dynamics of forced compression and structural collapse of this specific viral capsid provide a conceptual framework for describing the unique physico-chemical, thermodynamic and material properties of other virus particles.

## MATHERIALS AND METHODS

**Protein preparation:** Purified capsid preparations of empty CCMV particles were obtained using the purification procedures described elsewhere (41). Briefly the procedure consists of isolation of CCMV particles from cowpea plants 13 days after infection (42,43). After UV/Vis spectroscopy characterization, Fast Protein Liquid Chromatography (FPLC) and running a sample on SDS-PAGE to determine the size of the monomers, the virions were examined by transmission electron microscopy. These measurements revealed that the CCMV particles had the expected size of ~28 nm in diameter. The buffer conditions for the imaging and nanoindentation experiments were 50 mM sodium acetate and 1 M sodium chloride (pH = 5.0).

**Atomic Force Microscopy:** Hydrophobic glass slides were treated with an alkylsilane (4). The viral samples were kept under liquid conditions at all times; all the experiments were performed at room temperature. Capsid solutions were incubated for ~30 minutes on the hydrophobic glass slides prior to the indentation experiments. Olympus OMCL-RC800PSA rectangular, silicon nitride cantilevers (nominal tip radius < 20 nm and spring constant = 0.05 N/m) were calibrated in air yielding a spring constant of $\kappa = 0.0524 \pm 0.002$ N/m. Viral imaging (44,45) and nanoindentation (3) were performed on a Nanotec Electronica AFM (Tres Cantos, Spain). Experimental nanoindentation measurements were carried out using the cantilever velocity $v_f = 0.6$ and 6.0 μm/s. Additional measurements were performed at $v_f = 6.0 \times 10^{-2}$ and $6.0 \times 10^{-3}$ μm/s. The indentation data were analyzed using a home-written Labview program (National Instruments) as described previously (38).

**Self-Organized Polymer (SOP) Model:** In the topology-based SOP model, each residue is described by a single interaction center (C$_\alpha$-atom). The potential energy function of the protein conformation $U_{SOP}$ specified in terms of the coordinates $\{r_i\} = r_1, r_2, ..., r_N$ ($N$ is the total number of residues) is given by $U_{SOP} = U_{FENE} + U_{NB}^{ATT} + U_{NB}^{REP}$. The finite extensible nonlinear elastic (FENE) potential $U_{FENE} = -\sum_{i=1}^{N-1} \frac{k}{2} R_0^2 \log \left[ 1 - \frac{(r_{i,i+1} - r_{i,i+1}^0)^2}{R_0^2} \right]$ with the spring constant $k = 14$ N/m and the tolerance in the change of a covalent bond distance $R_0 = 2$ Å describes the backbone chain connectivity. The distance between residues $i$ and $i+1$, is $r_{i,i+1}$, and $r^0_{i,i+1}$ is its value in the native structure. We used the Lennard-Jones potential $U^{ATT}_{NB} = \sum_{i,j=i+3} \varepsilon_n \left[ (r^0_{ij}/r_{ij})^{12} - 2(r^0_{ij}/r_{ij})^6 \right] \Delta_{ij}$ to account for the non-covalent interactions that stabilize the native folded state. Here, the summation is performed over all the native contacts in the PDB structure; we assumed that if the non-covalently linked residues $i$ and $j$ ($|i\text{-}j| > 2$) are within the cut-off distance $R_C = 8$ Å in the native state, then $\Delta_{ij} = 1$, and is zero otherwise ($R_C = 8$ Å is the cut-off distance for calculating the non-bonded forces). The value of $\varepsilon_n$ quantifies the strength of the non-bonded interactions. We used $\varepsilon_n = \varepsilon_{inter} = 1.29$ kcal/mol and $\varepsilon_n = \varepsilon_{intra} = 1.05$ kcal/mol for the inter-chain contacts and intra-chain contacts (model parameterization is described in SM). The non-native interactions in the potential $U^{REP}_{NB} = \sum_{i,j=i+2} \varepsilon_r (\sigma_r/r_{ij})^6 + \sum_{i,j=i+3} \varepsilon_r (\sigma_r/r_{ij})^6 (1 - \Delta_{ij})$ were treated as repulsive. Here, the summation is performed over all the non-native contacts with the distance $> R_C$. A constraint is imposed on the bond angle formed by residues $i$, $i+1$, and $i+2$ by including the repulsive potential with parameters $\varepsilon_r = 1$ kcal/mol and $\sigma_r = 3.8$ Å, which determine the strength and the range of the repulsion. Six independent runs have been carried out for each simulation setup.

**Forced indentation simulations:** *Dynamic force measurements* *in silico* were performed using the SOP model and Langevin simulations accelerated on a GPU. We used the CCMV virus capsid, empty of RNA molecules (Protein Data Bank (PDB) code 1CWP) (40), and a spherical tip of radius $R_{tip} = 5$, 10, and 20 nm to compress the capsid along the two-, three-, and five-fold symmetry axis. The tip-capsid interactions were modeled by the repulsive Lennard-Jones potential, $V(r_i) = \varepsilon[\sigma/(r_i - R_{tip})]^6$, where $r_i$ are the $i$th particle coordinates, $\varepsilon = 4.18$ kJ/mol, and $\sigma = 1.0$ Å. We constrained the bottom portion of the CCMV by fixing five $C_\alpha$-atoms along the perimeter. The tip exerted the time-dependent force $\boldsymbol{f}(t) = f(t)\boldsymbol{n}$ in the direction $\boldsymbol{n}$ perpendicular to the surface of CCMV shell. The force magnitude $f(t) = r_f t$ increased linearly in time $t$ (force-ramp) with the force-loading rate $r_f = \kappa v_f$ ($v_f$ is the cantilever base velocity and $\kappa$ is the cantilever spring constant). In the simulations of "forward indentation", the spherical tip was moving towards the capsid with $v_f = 1.0$ μm/s and $v_f = 25$ μm/s ($\kappa = 0.05$ N/m). In the simulations of force-quenched retraction for $v_f = 1.0$ μm/s, we reversed the direction of tip motion. In simulations, we control the piezo displacement $Z$ (cantilever base), and the cantilever tip position $X$. The cantilever base (virtual particle) moves with constant velocity ($v_f$), which ramps up a force $f(t)$ applied to the capsid through the cantilever tip with the force-loading rate $r_f$. The resisting force $F$ from the capsid, which corresponds to the experimentally measured indentation force, can be calculated using the energy output. The capsid spring constant $k_{cap}$ can be obtained from an *FX* curve by calculating the slope $k_{cap} = dF/dX$.

**Analysis of simulation output:** *Analysis of structures:* To measure the extent of structural similarity between a given conformation and a reference state, we used the structure overlap $\chi(t) = (2N(N-1))^{-1} \sum \Theta(|r_{ij}(t) - r^0_{ij}| - \beta r^0_{ij})$. In $\Theta(x)$, Heaviside step function, $r_{ij}(t)$ and $r^0_{ij}$ are the inter-particle distances between the $i$-th and $j$-th residues in the transient and native structure, respectively ($\beta = 0.2$ is the tolerance for the distance change). *Analysis of energies:* We analyzed the potential energy $U_{SOP}$, and utilized Umbrella Sampling simulations (see SM; 46,47) to estimate the Gibbs energy ($\Delta G$), enthalpy ($\Delta H$), and entropy ($\Delta S$). *Normal Mode Analysis* was used to characterize the equilibrium vibrations (see SM) (48). We calculated the Hessian matrix for centers of mass of amino acid residues ($H_{IJ}$). The eigenvalues $\{\lambda_I\}$, and eigenvectors $\{R_I\}$ obtained numerically were used to calculate the spectrum of normal frequencies $\omega_i \propto \sqrt{\lambda_i}$ and normal modes $Q_i = \sum R_{ij} q_i$, ($q_i$—center-of-mass positions). In the *Essential Dynamics* approach (48), implemented in GROMACS (49), collective modes of motion describing the non-equilibrium displacements of amino acids are projected along the direction of global transition $X$ (indentation depth), characterized by the displacements $\Delta X(t) = X(t) - X_0$ from equilibrium $X_0$ (see SM). We diagonalized the covariance matrix $C(t) = \Delta X(t) \, \Delta X(t)^T = T\Lambda T^T$ to compute the matrix of eigenvalues $\Lambda$ and the matrix of eigenvectors $T$. These were used to find the projections $\Delta X(t)$ on each eigenvector $t_l$, $P_l(t) = t_l \Delta X(t)$.

## RESULTS

**AFM Indentation Experiments:** Before nanoindentation, AFM images of the capsid were recorded as depicted in Fig. 1A. Next, nanoindentation measurements were performed on the center of the CCMV capsid particle, and the corresponding force ($F$)–indentation ($Z$) curves (FZ curves) were recorded. The FZ curves quantify the mechanical response of the capsid (indentation force $F$) as a function of the piezo displacement (reaction coordinate $Z$). The FZ curves (Fig. 2) revealed that mechanical nanoindentation is a complex stochastic process, which might occur in a single step ("all-or-none" transition with a single force peak) or through multiple steps (several force peaks).

To characterize the experimental FZ curves, we focused on the common features, which include an initial linear-like indentation behavior followed by a sharp drop in force (Fig. 2). Next, we performed a fit of a straight line to the initial region of each FZ curve to determine the capsid spring constant $k_{cap}$, which quantifies the elastic compliance of the capsid, using the relationship $1/K = 1/\kappa + 1/k_{cap}$ for the cantilever plus capsid setup. Here, $K$ is the slope in the FZ curve (Fig. 2), and $\kappa$ is the cantilever spring constant (see Materials and Methods). We found that the average spring constant of CCMV is $k_{cap} = 0.17$ N/m at a loading rate $v_f = 0.6$ µm/s and $k_{cap} = 0.14$ N/m at $v_f = 6.0$ µm/s (Table I). Additional experiments showed that $k_{cap}$ does not change much over four decades of $v_f$ (Fig. 3A) (38).

The indentation force, where the linear-like regime in the FZ curve ends, corresponds to the critical force at which the mechanical failure of the capsid occurs. We analyzed the critical forces ($F^*$) by extracting peak forces observed in the FZ curves, and the corresponding transition distances ($Z^*$). The average critical force was determined to be $F^* = 0.71$ and $0.72$ nN for $v_f = 0.6$ µm/s and $v_f = 6.0$ µm/s, respectively, showing that critical force is not affected by loading rate in the regime of $v_f$ used (38). These experiments also showed a good agreement with previously published results (15), where it was found that $k_{cap} \approx 0.15$ N/m and $F^* \approx 0.6$ nN for comparable loading rates (0.02 µm/s − 2 µm/s range).

Next, we considered whether mechanical deformation was reversible. We performed measurements for small forward indentation followed by backward movement of the AFM tip, which we refer to as "force-quenched retraction". An example of such measurements clearly shows that there is a considerable difference between the mechanical response of the CCMV particle observed for small and large deformations (Fig. 3B). Whereas for large piezo displacements ($Z > 35$ nm) the deformation was irreversible with large hysteresis, for small displacements ($Z < 10$ nm) the deformation was completely reversible with almost no hysteresis. These results also agree with our previous findings (15).

**Forced Indentation *in silico*:** We performed indentation simulations using a spherical tip of radius $R_{tip}$ = 20 nm. The 250-fold computational acceleration on a GPU has enabled us to use experimentally relevant $v_f$ = 0.5 and 1.0 μm/s ($\kappa$ = 0.05 N/m) and span the experimental 0.1–0.2 s timescale. To provide the basis for comparison of the results of experiments and simulations, we analyzed the indentation force $F$ as a function of the piezo displacement $Z$ (FZ curves; see Materials and Methods). The theoretical FZ curves (Fig. 2C and 2D) compare well with the experimental FZ profiles (Fig. 2A and 2B). The simulated FZ curves also exhibit single-step and multi-step transitions. The average values of $k_{cap}$, $F^*$, and $Z^*$ compare well with their experimental counterparts (Table I). We also analyzed the dependence of $k_{cap}$ on $v_f$ (including additionally $v_f$ = 5 and 25 μm/s) and found that as in experiment, $k_{cap}$ was insensitive to the variation of $v_f$ (Fig. 3A). Next, we performed simulations of force-quenched retraction, where we used the CCMV structures generated in the forward deformation runs for $Z$ = 15, 25, 28, and 32 nm as initial conditions. Our simulations (Fig. 3C) agreed with experiments (Fig. 3B) in that indentation is fully reversible for small $Z$ = 15 nm displacement but irreversible beyond the critical distance, $Z$ > 25 nm.

To summarize, we have obtained an almost quantitative agreement between the results of dynamic force measurements *in vitro* and *in silico* (Fig. 2), including our findings regarding the weak sensitivity of the capsid spring constant $k_{cap}$ on changes in the cantilever velocity $v_f$ (Fig. 3A) and our results of force-quenched retraction (Fig. 3B and 3C). Hence, the SOP model of CCMV provides an accurate description of the capsid mechanical properties, which validates our approach. The good agreement between the results of experiments and simulations allowed us to probe features of CCMV shell that are not accessible experimentally.

**Mechanical properties of CCMV depend on local symmetry:** Next, we performed simulations of indentation under slow ($v_f$ = 1.0 μm/s) and fast ($v_f$ = 25 μm/s) force loading ($R_{tip}$ = 20 nm). The capsid was indented at different points on its surface: at the symmetry axes of the hexamer capsomeres (three-fold symmetry), the pentamer capsomeres (five-fold symmetry), and at the interface between two hexamers (two-fold symmetry) as described in Fig. S1 in the Supporting Material (SM). The FX profiles for the two-fold symmetry axes are displayed in Fig. 4. The FX curves for three- and five-fold symmetry are presented in Fig. S2 and S3, respectively. In agreement with experiment (Fig. 2B), the mechanical reaction of the capsid is elastic up to $X \approx$ 3-5 nm ($Z \approx$ 8-10 nm; linear regime) and quasi-elastic up to $X \approx$ 8-11 nm ($Z \approx$ 22 - 25 nm; linear-like regime), regardless of the capsid orientation (panel A in Fig. 4, S2, and S3). The FX curves generated under fast force loading ($v_f$ = 25 μm/s) showed a slightly steeper slope. Fitting a straight line to the initial portion of FX curves ($X$ < 3 nm) taken at $v_f$ = 1.0 μm/s yielded the spring constant $k_{cap} \approx$ 0.11 N/m, 0.10 N/m, and 0.12 N/m for two-, three-, and five-fold symmetry, respectively (Table II).

A more detailed analysis showed that $k_{cap}$ is significantly varying during simulated particle deformation: $k_{cap}$ varies from 0.06 - 0.14 N/m, 0.05 - 0.10 N/m, and 0.04 - 0.12 N/m for the two-, three-, and five-fold symmetry axes, respectively in the initial deformation regime. Fluctuations in $k_{cap}$ show systematic differences for the icosahedral symmetry axes (Fig. 4, S2, and S3). When probing along the two- and five-fold axes, the curves of $k_{cap}$ versus $X$ show two maxima: the first maximum is at $X \approx$ 2-3 nm (for two- and five-fold symmetry), and the second maximum is at $X \approx$ 5-6 nm (two-fold symmetry) and 11-12 nm (five-fold symmetry). For the three-fold symmetry, $k_{cap}$ shows one broad skewed peak centered at $X \approx$ 5 nm. To explain this effect, we performed structural analysis of CCMV conformations, and found that at various stages of mechanical deformation different types (pentamers and hexamers) and number of capsomers cooperate to withstand the mechanical stress (Fig. S1).

The collapse transition occurs in the 11 - 15 nm range (Fig. 4, S2, and S3). For $v_f = 1.0$ µm/s, the average critical forces from experiments and simulations agreed ($F^* = 0.71$ nN; see Fig. 2 and Tables I and II). Here, the bottom portion of the shell becomes increasingly more flat (see snapshots in panel C in Fig. 4, S2, and S3), and the capsid undergoes a spontaneous shape change from a roughly spherical state to a non-spherical collapsed state, which is reflected in the sudden force drop and decrease of $k_{cap}$ to zero (panel B in Fig. 4, S2, and S3). Under fast loading ($v_f = 25$ µm/s), force peaks were not detected (panel A in Fig. 4, S2, and S3). To quantify the extent of the structural collapse, we monitored the structure overlap $\xi$. In the transition regime, $\xi$ decreased from $\xi = 1$ (native state) to $\xi = 0.65$ (collapsed state) for all symmetry types (panel C in Fig. 4, S2, and S3). Hence, notwithstanding the large scale transition, the capsid structure remained 65% similar to the native state. For faster $v_f = 25$ µm/s, $\xi$ decreased by <10%, indicating that fast force loading leaves the local arrangements of capsomers unaffected. At $X >$ 15 nm, CCMV entered the post-collapse, second linear-like regime (panels A and B in Fig. 4, S2, and S3). Here, as the tip approached the solid surface, $k_{cap}$ increased sharply.

Next, we reversed the direction of tip motion using the structures for the collapsed state obtained for $Z = 15, 25, 28$, and 32 nm ($X = 5, 11, 15$, and 19 nm) as initial conditions. In agreement with AFM data (Fig. 3B), the theoretical force-retraction curves for $v_f = 1.0$ µm/s showed that the mechanical compression of CCMV was fully reversible in the elastic regime for $X = 5$ nm (no hysteresis), nearly reversible in the quasi-elastic regime for $X = 11$ nm (small hysteresis), but irreversible after the transition had occurred ($X = 15$ and 19 nm; panel A in Fig. 4, S2, and S3). We also analyzed the progress of CCMV shell restructuring by monitoring $\xi$ as a function of time (panel C in Fig. 4, S2, and S3), and found that the CCMV shell recovered its original shape ($\xi = 1$) in the millisecond timescale.

**Thermodynamics of CCMV indentation:** We evaluated the total work of indentation $w$ by integrating the area under the FX curves. We repeated this procedure for the retraction curves to evaluate the reversible work $w_{rev}$. Estimation of the relative difference $(w - w_{rev})/w$ showed that in the elastic and quasi-elastic regime ($X < 11$ nm; $Z < 25$ nm) ~12% of $w$ was dissipated. This agrees with the experimental finding that the fraction of energy returned upon retraction is ~90% (15). In the transition range (11 nm $\leq X \leq$ 15 nm; 25 nm $\leq Z \leq$ 30 nm), where the retraction curves showed large hysteresis especially for the three-fold symmetry, $(w - w_{rev})/w \approx 75\%$.

Because $w_{rev}$ is equal to the Gibbs energy change, i.e., $w_{rev} = \Delta G = \Delta H - T\Delta S$, where $\Delta H$ and $\Delta S$ are the enthalpy change and entropy change, we estimated $\Delta H$ and $T\Delta S$. The results for $\Delta H$ and $T\Delta S$ for indentation along the two-, three-, and five-fold symmetry axes ($R_{tip} = 20$ nm; $v_f = 1.0$ µm/s) are displayed, respectively, in Fig. 4, S2, and S3. In the linear regime and linear-like regime (X < 10-11 nm), $\Delta H$ and $T\Delta S$ display a parabolic dependence on $X$ and $\Delta H \approx T\Delta S$; in the (11-15 nm) transition range and in the post-collapse regime (X > 15 nm), $\Delta H$ and $T\Delta S$ level off, each attaining a plateau, and $\Delta H > T\Delta S$. The dependence of $\Delta H$ on $X$ under fast force-loading conditions is more monotonic. The curves of $\Delta G$, $\Delta H$, and $T\Delta S$ attain some constant values $\Delta G_{ind}$, $\Delta H_{ind}$, and $T\Delta S_{ind}$ at $X = 20$ nm, which correspond to the Gibbs energy, enthalpy, and entropy of indentation (Table II). We also mapped the equilibrium energy landscape $\Delta G$ using the Umbrella Sampling simulations (see Experimental Procedures) and resolved the profiles of $\Delta H$ and $T\Delta S$ (the *inset* in panel D in Fig. 4, S2, and S3). The equilibrium values of $\Delta G_{ind}$, $\Delta H_{ind}$, and $T\Delta S_{ind}$ are accumulated in Table II. The thermodynamic functions indicate that mechanical compression of the CCMV shell requires a considerable investment of energy, and that $\Delta G_{ind}$, $\Delta H_{ind}$, and $T\Delta S_{ind}$ vary with the local symmetry under the tip.

**Mechanical response of CCMV depends on geometry of force application:** We performed simulations of the nanoindentation of CCMV along the two-fold symmetry axis using a tip of smaller radius $R_{tip} = 10$ and 5 nm. The FX profiles, spring constant $k_{cap}$, structure overlap $\xi$, and thermodynamic functions obtained for $R_{tip} = 10$ nm (Fig. S4) can be compared with the same quantities obtained for $R_{tip} = 20$ nm (Fig. 4). We present our findings for $R_{tip} = 10$ nm; results for $R_{tip} = 5$ nm show a similar tendency (data not shown). The FX curves for $v_f = 1.0$ (and 25 μm/s) shows a less steep $k_{cap}$, the collapse transition is less pronounced and starts sooner ($X^* \approx$ 9 nm), and the critical force is lower ($F^* \approx 0.6$ nN) for indentation with a smaller tip (panels A and B in Fig. S4). The $k_{cap}$ versus $X$ dependence shows two peaks, but the second peak at $X \approx 6$ nm is weaker (compared to the results for $R_{tip} = 20$ nm). The overlap $\xi$ decreased to 0.75, implying that in the collapsed state the CCMV shell remained $\approx 75\%$ similar to the native state (panel C in Fig. S4). $\Delta H$ and $T\Delta S$ show a familiar parabolic dependence on $X$ (as for $R_{tip} = 20$ nm), but these level off at somewhat lower values (panel D in Fig. S4). Numerical estimates of $k_{cap}$, $\Delta G_{ind}$, $\Delta H_{ind}$, and $T\Delta S_{ind}$ obtained for $R_{tip} = 10$ and 5 nm (Table III) are directly proportional to the tip size as $k_{cap}$, $\Delta G_{ind}$, $\Delta H_{ind}$, and $\Delta S_{ind}$ all decrease with $R_{tip}$. We also performed simulations using 10 and 5 nm tip but for the three- and five-fold symmetry axes, and arrived at the same conclusions (data not shown).

**Equilibrium and non-equilibrium dynamics of CCMV:** To resolve the dynamic determinants underlying the mechanical response of CCMV, we first analyzed equilibrium properties of the capsid using Normal Mode Analysis (see Material and Methods and SM). We calculated the spectra of normal modes for $C_\alpha$-atoms for a single hexamer, single pentamer, and for the whole CCMV particle (Fig. S5). Because the spectra for a pentamer and hexamer were identical, we only display the spectrum for a hexamer, which practically overlaps with the spectrum for the CCMV shell, implying that normal displacements of the CCMV shell and its constituents are similar, especially in the 50-500 cm$^{-1}$-range of frequencies. Analysis of structures revealed that the more global modes of motion in the low-frequency part of the spectrum ($\leq 50$ cm$^{-1}$) involve the "out-of-plane" expansion-contraction excitations and the "in-plane" concerted displacements of capsomers. The more local modes (100-250 cm$^{-1}$ range) are small-amplitude displacements of the secondary structure elements. The high frequency 300-450 cm$^{-1}$ end of the spectrum is dominated by the local vibrations of amino acids (Fig. S5).

To describe the far-from-equilibrium mechanical response of CCMV, we employed the Essential Dynamics approach (see Materials and Methods and SM), which helps to single out the most important types of motion, showing the largest contribution in the direction of global transition (indentation depth $X$). Importantly, the essential dynamics modes should not be confused with the equilibrium normal modes or instantaneous normal modes. We examined separately the elastic regime ($X \leq 5$ nm) and the transition regime (11 nm $\leq X \leq 15$ nm) using the simulation output for the two-fold symmetry (Fig. 4). We resolved the principal coordinates $P_I(t)$, collective variables describing the non-equilibrium dynamics of the system, and analyzed the relative displacement for each $I$-th mode given by the ratio $\left\langle (x_I - x_{I0})^2 \right\rangle / \sum \left\langle (x_I - x_{I0})^2 \right\rangle$ of the average squared displacement $\left\langle (x_I - x_{I0})^2 \right\rangle$ to the total squared displacement $\sum \left\langle (x_I - x_{I0})^2 \right\rangle$. It turns out that the first two modes account for ~85% of dynamics of CCMV ("essential subspace"); the remaining modes are negligible in terms of the displacement amplitude.

The CCMV dynamics in the essential subspace is displayed in Fig. 5. We characterized large-amplitude non-equilibrium displacements, which contribute the most to the mechanical compression of CCMV. In the elastic regime, the first mode (mode 1: 77% of dynamics) corresponds to the large-amplitude out-of-plane compression, which results in the capsid bending. Here, the top and bottom portions of the capsid become flat, while the capsid sides expand outward (Fig. 5A). The second mode (mode 2: 8% of dynamics) represents direct coupling of the in-plane displacements of capsomers and the out-of-plane capsid bending, for which the in-plane displacements and the out-of-plane capsid bending occur at the same time. Here we observe that the arrangement of capsomers on the spherical surface change from the more ordered to the less ordered, and the capsid structure loses its near-spherical symmetry (Fig. 5A). In the transition regime, the in-plane and out-of-plane displacements are strongly coupled. That is, these displacements are neither purely in-plane nor out-of-plane excitations. The first mode (mode 1) represents the collapse transition, which is accompanied by the lateral translocation of the capsomers towards the tip-surface contact area (Fig. 5B). The second mode (mode 2) is dominated by the lateral translocation and twisting motions of hexamers and pentamers in the clock-wise and counter clock-wise direction around their symmetry axes, respectively (Fig. 5B).

## DISCUSSION

By coupling dynamic force measurements *in vitro* and *in silico*, here we have directly compared, for the first time, the experimental data with simulation data for CCMV shell obtained under identical conditions of the mechanical force load. Excellent agreement between the experimental and simulation results validated our theoretical approach. Larger variation in the experimental FZ profiles are due to the fact that in experiments not only three different icosahedral orientations were probed (two-, three-, and five-fold symmetry) by the AFM tip, but also various intermediate orientations. This undertaking had enabled us to interpret the experimental forced indentation patterns in unprecedented detail with regards to the structural and thermodynamic changes in the CCMV capsid in response to external mechanical deformation.

The main results are: 1) The physical properties of the CCMV shell are dynamic but local characteristics of the structure, and the mechanical response of the capsid depends not only on the symmetry of the local capsomer arrangement under the tip, but also on the indentation depth. 2) The mechanical characteristics of CCMV - the critical force and transition distance - depend on how rapidly the compressive force is increased. 3) The physical properties of the CCMV particle depend on the geometry of mechanical perturbation, as the mechanical response changes with tip size. 4) The extent to which the mechanical deformation of the CCMV shell can be retraced back reversibly depends on the indentation depth. 5) In the elastic regime of deformation, the "out-of-plane" excitations dominate the near-equilibrium displacements of capsomers, but these "in-plane" modes are strongly coupled in the far-from-equilibrium transition range. 6) The entropic and enthalpic contributions are almost equally important for the capsid stiffening, whereas the capsid softening and transition to the collapsed state is driven mainly by the enthalpy change.

Our conclusion about the local nature of physical properties also fits with previous modeling of Hepatitis B Virus, which showed that permanent deformation of the shell was due to local rearrangements of the capsid proteins (21). That the CCMV shell displays multiple modes of mechanical resistance, which depends on the indentation depth, agrees well with recent studies, which showed two dynamic regimes to be responsible for the CCMV capsid stiffening and softening (34). The existence of multiple modes is reflected in the non-monotonic dependence and maxima of $k_{cap}$ as a function of $X$. The periods of mechanical resistance (stiffening), during which an increasingly larger portion of protein chains find themselves in the tip-shell contact area, are interrupted by the periods when the capsid yields to force (softening). The first peak of $k_{cap}$ at $X \approx 2$-3 nm (Fig. 4 and S3), agrees with the previous results from finite element analysis (34). The second peak at $X \approx 6$ nm and $X \approx 11$ nm for the two-fold and five-fold symmetry described here correspond to the capsid softening beyond $X \approx 10$ nm (34).

The weak dependence of CCMV mechanical properties on the rate of change of compressive force is not unexpected, because the positions of transition states and barrier heights on the energy landscape depend upon how force is applied to the system. Under fast force loading ($v_f = 25$ μm/s), the energy pumped into the system is more than sufficient to overcome the energy barrier, and the transition to the collapsed state is not well-pronounced (no force-drop in the FX curves in Fig. 4, S2, and S3). The force peaks are observed under slow loading ($v_f = 0.5 - 6$ μm/s) because the amount of energy is comparable to the energy barrier for the collapse transition. The FZ curves for the CCMV capsid obtained using the finite element analysis (34) agree with our result for $v_f = 25$ μm/s. This indicates that conditions of force application used in the finite element analysis correspond to the fast indentation case. The FZ profiles from the finite element analysis and our own results also agree in that, under fast force loading, differences in the mechanical response of CCMV for different symmetries disappear.

In single-molecule manipulation on virus particles, mechanical force requires a physical contact between a system and a probe. Hence, their shape, size difference, and the direction of force become important factors. When a virus is indented by a plane ($R_{tip} >> R$ – radius of a virus shell), all residues in the tip-capsid contact area are pushed in the same direction; when a virus is indented by a small sphere ($R_{tip} \approx R$), different residues are displaced in different directions. Our results show that the mechanical characteristics - FX profile, spring constant, critical force, and indentation depth (Fig. S4) - all change with probe size, and that $\Delta G_{ind}$, $\Delta H_{ind}$, and $T\Delta S_{ind}$ are directly proportional to $R_{tip}$ (Table III). A smaller tip means a smaller tip-capsid contact area, and, hence, weaker mechanical response and lower associated energy costs.

In the elastic regime, quasi-elastic regime, and transition regime (Fig. 4, S2, and S3), the deformation is reversible for short $X$ and almost reversible for longer $X$. In the post-collapse regime, the mechanical compression is irreversible. In fact, these same findings can be rationalized using our results from Essential Dynamics (Fig. 5). In the elastic regime and quasi-elastic regime, the first mode is dominated by the out-of-plane displacements of pentamers and hexamers. Hence, when a compressive force is quenched, as in the retraction experiments, the first mode provides a mechanism for capsid reshaping, and the amount of energy dissipated is small. In the transition range, the two most important modes represent strongly coupled out-of-plane and in-plane displacements. Here, the capsid is capable of restoring its original shape, but capsid restructuring comes at a cost of exciting additional degrees of freedom and, hence, a larger amount of dissipated energy.

The question exists whether the property of a whole system can be represented by a sum of the properties of its structural elements (50). For the CCMV capsid dynamics at equilibrium, our results from Normal Mode analysis (37,51) provide the affirmative answer. The spectra of eigenmodes for an isolated single pentamer or hexamer and for the whole capsid show only small differences at low frequencies (<50 cm$^{-1}$), due to icosahedrally symmetric modes present in the full capsid, but practically overlap with that for the whole shell in the 50 − 500 cm$^{-1}$-range (Fig. S5). This result is not surprising. The differences in the small-amplitude equilibrium fluctuations of residue positions for local modes are negligible for the penton, hexon, and full capsid. Of course, these modes represent collective motions, which correspond to penton, hexon, and full capsid decompositions; yet, when compared at the whole shell level, these motions in the penton and hexon units, and in the full capsid are nearly identical (Fig. S5). Hence, domain and capsomer interactions have little effect on equilibrium properties of CCMV.

Under non-equilibrium conditions of mechanical deformation, the near-spherical symmetry of the capsid is broken and different capsomers start playing different roles. In this regime, we can no longer use a concept of equilibrium normal modes. We employed the Essential Dynamics approach to characterize large-amplitude displacements of capsomers under far-from-equilibrium conditions. Although in the linear regime the main mode of collective motions is dominated by the "out-of-plane" displacements, there are no pure "out-of-plane" and "in-plane" modes either in the elastic regime or in the transition range (Fig. 5). These coupled non-equilibrium essential modes of motion, which accompany the CCMV transition to the collapsed state, cannot be reconstructed using "a linear combination" of the out-of-plane modes and the in-plane modes. The concerted in-plane displacements mediate rearrangements of pentamers and hexamers on the CCMV surface, which leads to capsid stiffening reflected in the non-monotonic dependence of $k_{cap}$ (Figs. 4, S2, S3, and S4). These are exact results, i.e. we have arrived at these conclusions by analyzing the output from Essential Dynamics calculations. Similar findings have been reported by other research groups (52).

We mapped the energy landscape for the mechanical deformation of the CCMV capsid (Fig. 4, S2, and S3). The similarity of non-equilibrium estimates of $\Delta G_{ind}$, $\Delta H_{ind}$, and $\Delta S_{ind}$ (from FX curves) and their equilibrium counterparts (from Umbrella Sampling) implies that slow force loading ($v_f = 1.0$ μm/s) corresponds to near-equilibrium conditions of force application. Both the entropic and enthalpic contributions to $\Delta G$ (6.5-6.9 MJ/mol) are important: the entropy change $T\Delta S_{ind}$ (5.1-5.8 MJ/mol) is roughly half the enthalpy change $\Delta H_{ind}$ (11.5-12.8 MJ/mol) for all three symmetry types (Table II). There are variations in the values of $\Delta G_{ind}$, $\Delta H_{ind}$, and $T\Delta S_{ind}$ for different symmetries: these functions for five-fold symmetry differ by ~10% from the same functions for two- and three-fold symmetry (Table II). Hence, our findings stress the importance of any particular capsid's discrete nature and local protein subunit(s)/capsomer symmetry when virus shells are tested mechanically.

The potential energy of protein chains ($U_{SOP}$) sharply increases in the transition range where the capsid alters its shape from the convex to the concave (tip-indented convex down). These shape alterations are captured by the enthalpy change $\Delta H$ (Fig. 4). Compared to the elastic regime of CCMV deformation ($X < 3$-5 nm), where $\Delta H$ increases by ~3 MJ/mol, in the transition region (11 nm $< X < 15$ nm) $\Delta H$ increases three-fold to ~10 MJ/mol (Fig. 4D). Here, the large-amplitude out-of-plane displacements mediate the capsid bending inward. Hence, in the quasi-elastic regime before the collapse transition occurs the out-of-plane collective modes contribute mainly to the enthalpy change $\Delta H$. Although small-amplitude in-plane displacements are coupled to the out-of-plane modes, the main effect from in-plane displacements is concerted transitions - displacements, translocations, and twisting, from the more ordered to the less ordered phase formed by protein chains (Fig. 5). Hence, the in-plane modes contribute mainly to the entropy change *TΔS,* which increases two-fold from ~3 MJ/mol in the elastic regime to ~6 MJ/mol in the transition range (Fig. 4). The map of local potential energy for protein chains forming capsomers shows that there is more energy stored in pentamers than in hexamers in the elastic regime and in quasi-elastic regime (Fig. S6). This correlates well with the inhomogeneous stress distribution in CCMV capsid found earlier using other methods (53). However, this picture is more mixed in the transition range. Hence, under tension, the same protein chains forming capsomers play different roles in the energy distribution, which changes with the indentation depth.

When the capsid is undergoing the global transition to the collapsed state, the average structure of the protein chains forming capsomers is affected, but to a limited extent. This is reflected in the small decrease of the structure overlap to $\xi \approx 0.6$ for the slow force loading ($v_f = 1$ μm/s). Under fast loading ($v_f = 25$ μm/s), or for smaller tip ($R_{tip} = 10$ nm), the decrease in $\xi$ is even smaller (Fig. 4, S2, and S3, and S4). This stands in contrast to mechanical protein unfolding where transitioning to the globally unfolded state occurs concomitant with the disruption of native interactions stabilizing the tertiary and secondary structures of the native fold. Hence, in the context of mechanical deformation of a capsid, force-induced spontaneous shape changing does not imply substantial structural transitions on the local scale.

We have advanced a conceptual understanding of the unique physical properties of capsids - their dynamic and local structural features, and their dependence on the rate of change and geometry of external physical stimulus. We have resolved the origin of multiple modes of elastic compliance leading to mechanical stiffening and softening effects, and have characterized (ir)reversibility of the mechanical deformation of virions. We have described specific roles played by the in-plane and out-of plane non-equilibrium collective modes of the capsomers' displacements and their connection to the thermodynamic functions. Because these properties are likely to be shared among different virion classes, the results of these studies are important to understand the nanomechanics of other protein shells. Furthermore, profiling the structural, dynamic, and thermodynamic characteristics of capsids can illuminate different aspects of their biological properties and function. Also, biotechnological applications of protein nanocontainers range from catalysis in constrained environments in organic synthesis to transport and delivery of substrates into cells in nanomedicine, and to building blocks in nanotechnology (54,55). Our combined *in vitro* and *in silico* techniques may become a widely used approach to reveal the structure-dynamics relationship for biologically derived nanoparticles.

**Acknowledgements:** This work was supported by the Russian Ministry of Education and Science (Grant 14.A18.21.1239 to VB), by the "Physics of the genome" program grant from Fundamenteel Onderzoek der Materie (FOM) (to GJLW), and by the National Science Foundation (grant MCB-0845002 to RID).

**SUPPORTING CITATIONS:**

References (56,57) appear in the Supporting Material.

**Figure 1.** The CCMV capsid: Panel **A**: AFM image of a CCMV capsid ($z_{max}$ = 30 nm, scale bar = 20 nm). Panel **B**: Structure of CCMV (PDB entry: 1CWP); protein domains forming pentamers are colored in blue, while the same protein domains in hexamers are shown in red and orange.

**Figure 2.** Mechanical indentation of the CCMV capsid *in vitro* (panels **A** and **B**) and *in silico* along two-, three-, and five-fold symmetry axes (panels **C** and **D**). Show in different color are the most representative trajectories of forced indentation. The force (F)-distance (Z) profiles (FZ curves) obtained from experimental AFM measurements for $v_f$ = 0.6 and 6.0 µm/s are compared with the theoretical FZ curves obtained for $v_f$ = 0.5 and 1.0 µm/s ($R_{tip}$= 20 nm). The black dash-dotted control lines correspond to the cantilever deforming against the glass surface. The insets to panels **C** and **D** show the corresponding FX profiles. Due to the stochastic nature of the mechanical indentation, the values of critical force ($F^*$, force peaks), transition distance ($Z^*$) and indentation depth ($X^*$) are varying. The FZ curves with a single (several) force peak represent single-step (multi-step) indentation transitions.

**Figure 3.** Panel **A**: Log-linear plot of the spring constant of CCMV capsid $k_{cap}$ versus the cantilever velocity $v_f$. The experimental data are compared with the results of simulations. Panel **B** and **C**: Reversible and irreversible deformation of the CCMV capsid obtained experimentally for $v_f$ = 6.0 µm/s (panel **B**), and theoretically for $v_f$ = 1.0 µm/s (panel **C**). Deforming the capsid with a small force of ~0.3 nN (experiment) and ~0.5 nN (simulations) resulted in the reversible mechanical deformation with no hysteresis. Increasing the force beyond ~0.5 nN led to the irreversible deformation - the forward indentation (solid curves) and backward retraction (dotted curves) do not follow the same path (hysteresis). Color denotation is presented in the graphs.

**Figure 4.** Mechanical indentation *in silico* of CCMV capsid along the two-fold symmetry axis (see panel **D** and Fig. S1). Show in read and blue color are the two trajectories. The cantilever tip ($R_{tip}$= 20 nm) indents the capsid in the direction perpendicular to the capsid surface ($v_f$ = 1.0 µm/s). Results for the forward deformation and backward retraction are represented by the solid and dotted curves, respectively; results obtained for $v_f$ = 25 µm/s are shown for comparison (dashed black curve). Panel **A**: The FX curves. The grey line corresponds to the linear fit of initial elastic regime, i.e. $X$ < 3-5 nm. Panel **B**: Capsid spring constant, $k_{cap}$ versus $X$. Panel **C**: Structure overlap $\xi$ versus $X$: the *inset* shows the time-dependence of $\xi$ for the backward retraction, which quantifies the progress of capsid restructuring. Panel **D**: The enthalpy change $\Delta H$ and entropy change $T\Delta S$ from the FX curves generated for $v_f$ = 1.0 µm/s (the dashed curve of $\Delta H$ generated for $v_f$ = 25 µm/s is presented for comparison). The *inset* shows equilibrium energy change $\Delta G$ along the reaction coordinate $X$ from Umbrella Sampling calculations. Also shown are the CCMV capsid structures (top view and profile) for different extents of indentation. The tip-capsid surface contact area shown in red color (see also Fig. S1).

**Figure 5.** Dynamics of the CCMV shell in terms of the non-equilibrium displacement of pentamers (shown in blue) and hexamers (shown in red color). Displayed are the structures for the first two modes of the collective excitations (black arrows), projected along the reaction coordinate (large arrow) in the elastic regime (panel **A**) and transition regime (panel **B**) of mechanical compression. For each mode, the upper structure is the reference state. In the lower structure, we showed the type and amplitude of displacement by juxtaposing the representative conformation with the reference state (shown in gray color).

**Figure 6.** Surface map of the potential energy (color scale for $U_{SOP}$ is in the graph) for three representative structures of the CCMV shell (top view) observed in the course of deformation *in silico* at $Z = 3$ nm (panel **A**), 18 nm (panel **B**), and 27 nm (panel **C**). The direction of motion of the tip is perpendicular to the CCMV surface as indicated. The map shows a gradual increase in the potential energy of proteins forming pentamers and hexamers as changes to the global structure occur.

**Table I:** Mechanical properties of the CCMV capsid from indentation measurements *in vitro* and *in silico*: the average spring constant $k_{cap}$, average critical force $F^*$, and average transition distance $Z^*$. These quantities were calculated by averaging over all FZ curves (all symmetry types). Experimental measurements were performed using the cantilever velocity $v_f = 0.6$ μm/s and 6.0 μm/s; simulations were carried out using $v_f = 0.5$ μm/s and 1.0 μm/s (Fig. 2). The experimental results for $v_f = 6.0$ μm/s and simulation results for $v_f = 1.0$ μm/s are shown in parenthesis.

| Indentation | $k_{cap}$,N/m | $F^*$, nN | $Z^*$, nm |
|---|---|---|---|
| *in vitro* | 0.17±0.01(0.14±0.02) | 0.71±0.08(0.72±0.07) | 21.0±3.6(20.8±1.7) |
| *in silico* | 0.11±0.01(0.11±0.02) | 0.77±0.03(0.71±0.02) | 24.7±2.1(25.5±0.9) |

**Table II:** Mechanical and thermodynamic properties of the CCMV capsid from indentation measurements *in silico* performed along the two-fold, three-fold, and five-fold symmetry axes (Fig. S1): critical force $F^*$, indentation depth $X^*$, spring constant $k_{cap}$, and thermodynamic functions - Gibbs energy change $\Delta G$, enthalpy change $\Delta H$, and entropy change $T\Delta S$. Theoretical estimates of these quantities were obtained by averaging the results of 3 trajectories, using $R_{tip} = 20$ nm and $v_f = 1.0$ μm/s (see also Fig. 4, and Figs. S2 and S3). The values of $\Delta G$, $\Delta H$, and $T\Delta S$ correspond to the total change in these quantities observed at $X = 20$ nm (Z = 30 nm) indentation. The range of variation of $k_{cap}$ (from Figs. 4, S2, and S3) and the equilibrium estimates of $\Delta G$, $\Delta H$, and $T\Delta S$ (from Umbrella Sampling calculations) are shown in parentheses.

| Symmetry | $F^*$, nN | $X^*$, nm | $k_{cap}$, N/m | $\Delta G_{ind}$, MJ/mol | $\Delta H_{ind}$, MJ/mol | $T\Delta S_{ind}$, MJ/mol |
|---|---|---|---|---|---|---|
| two-fold | 0.71±0.02 | 9.1±1.0 | 0.11 (0.06-0.14) | 4.5(6.9) | 11.5(12.8) | 7.0 (5.8) |
| three-fold | 0.68±0.02 | 11.9±0.5 | 0.10 (0.05-0.10) | 5.1(6.8) | 11.7(12.6) | 6.6(5.8) |
| five-fold | 0.69±0.02 | 14.2±0.5 | 0.12 (0.04-0.12) | 4.1(6.5) | 12.5(11.5) | 8.4(5.1) |

**Table III:** Mechanical and thermodynamic properties of the CCMV capsid from indentation measurements *in silico* performed at $v_f = 1.0$ μm/s, along the two-fold symmetry axis (Fig. S1) - spring constant $k_{cap}$, and Gibbs energy change $\Delta G_{ind}$, enthalpy change $\Delta H_{ind}$, and entropy change $\Delta S_{ind}$ - are compared for the spherical tips of different radius $R_{tip} = 20$ nm, 10 nm, and 5 nm. The estimates of $k_{cap}$, $\Delta G_{ind}$, $\Delta H_{ind}$, and $T\Delta S_{ind}$ are obtained from a single FX curve for each different $R_{tip}$ and correspond to the total change in these quantities observed at $X = 20$ nm (Z = 30 nm) indentation. Simulation data for $R_{tip} = 20$ nm and 10 nm are shown in Fig. 4 and S4, respectively.

| $R_{tip}$, nm | $k_{cap}$, N/m | $\Delta G_{ind}$, MJ/mol | $\Delta H_{ind}$, MJ/mol | $T\Delta S_{ind}$, MJ/mol |
|---|---|---|---|---|
| 20 | 0.090 | 4.5 | 11.5 | 7.0 |

| 10 | 0.075 | 3.9 | 9.6 | 5.7 |
|----|-------|-----|-----|-----|
| 5  | 0.069 | 1.8 | 4.9 | 2.2 |

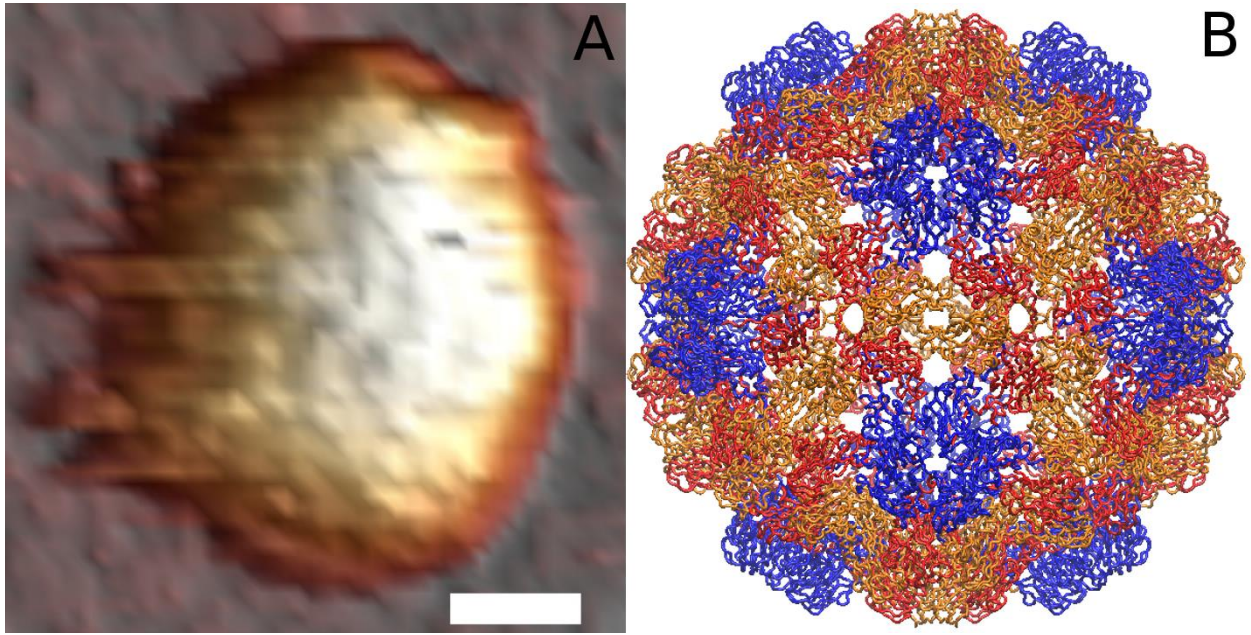

**Figure 1 (Kononova, Snijder, Brasch, Cornelissen, Dima, Marx, Wuite, Roos, Barsegov)**

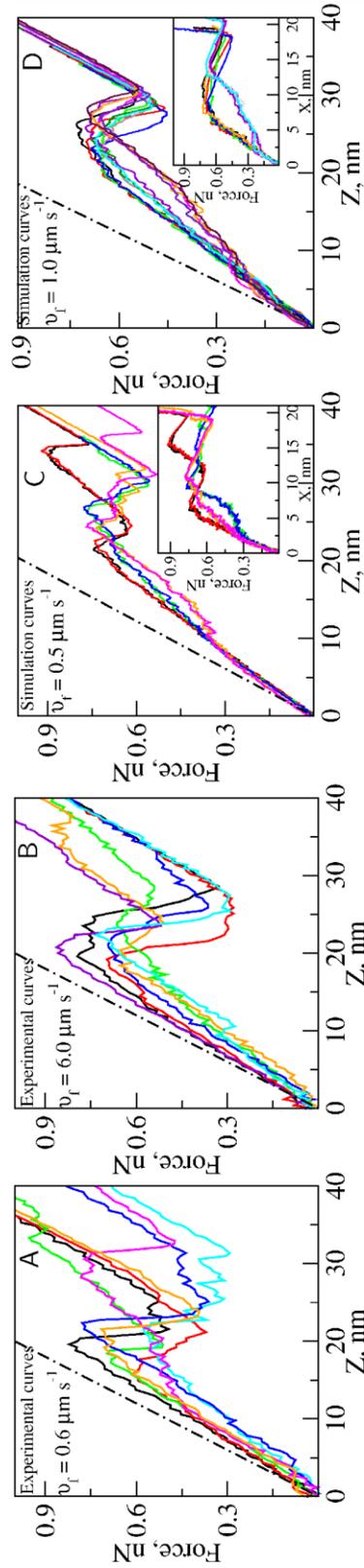

**Figure 2 (Kononova, Snijder, Brasch, Cornelissen, Dima, Marx, Wuite, Roos, Barsegov)**

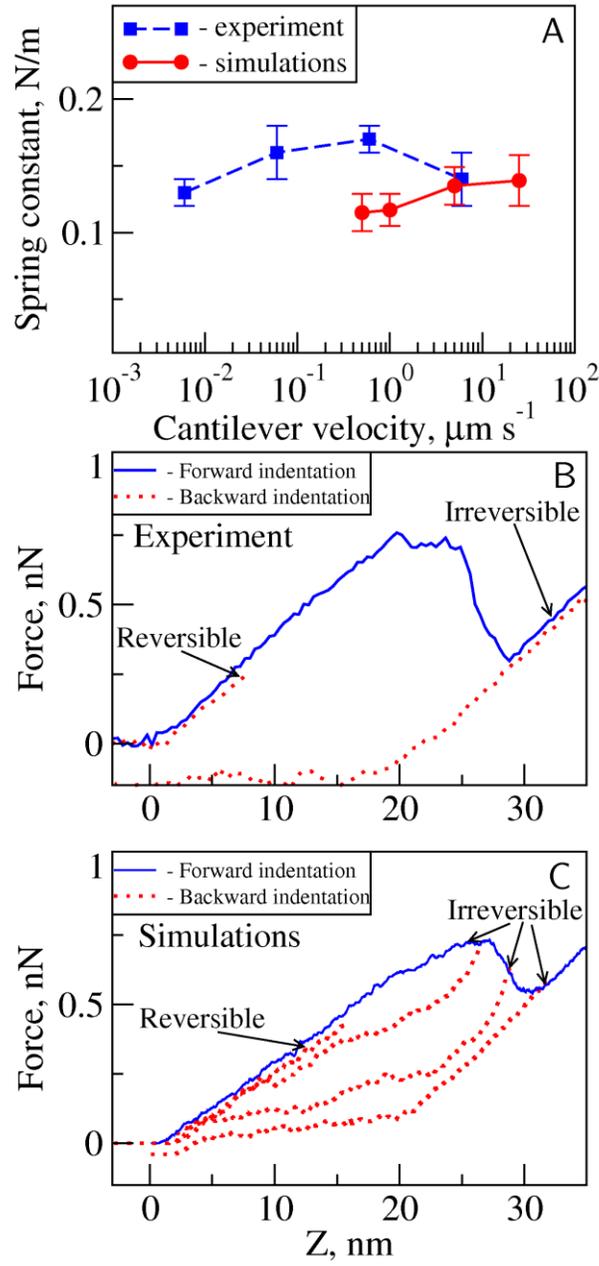

**Figure 3 (Kononova, Snijder, Brasch, Cornelissen, Dima, Marx, Wuite, Roos, Barsegov)**

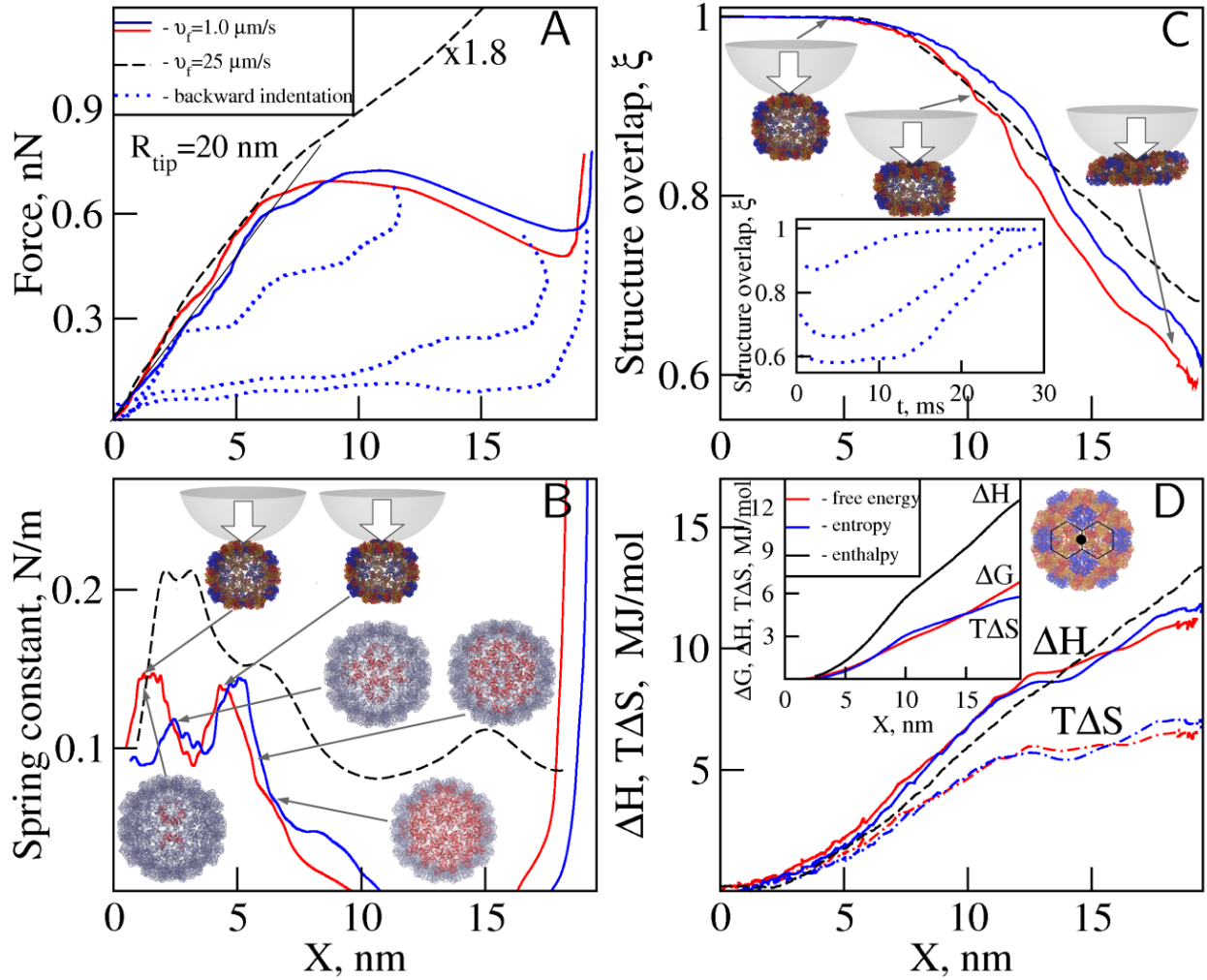

**Figure 4 (Kononova, Snijder, Brasch, Cornelissen, Dima, Marx, Wuite, Roos, Barsegov)**

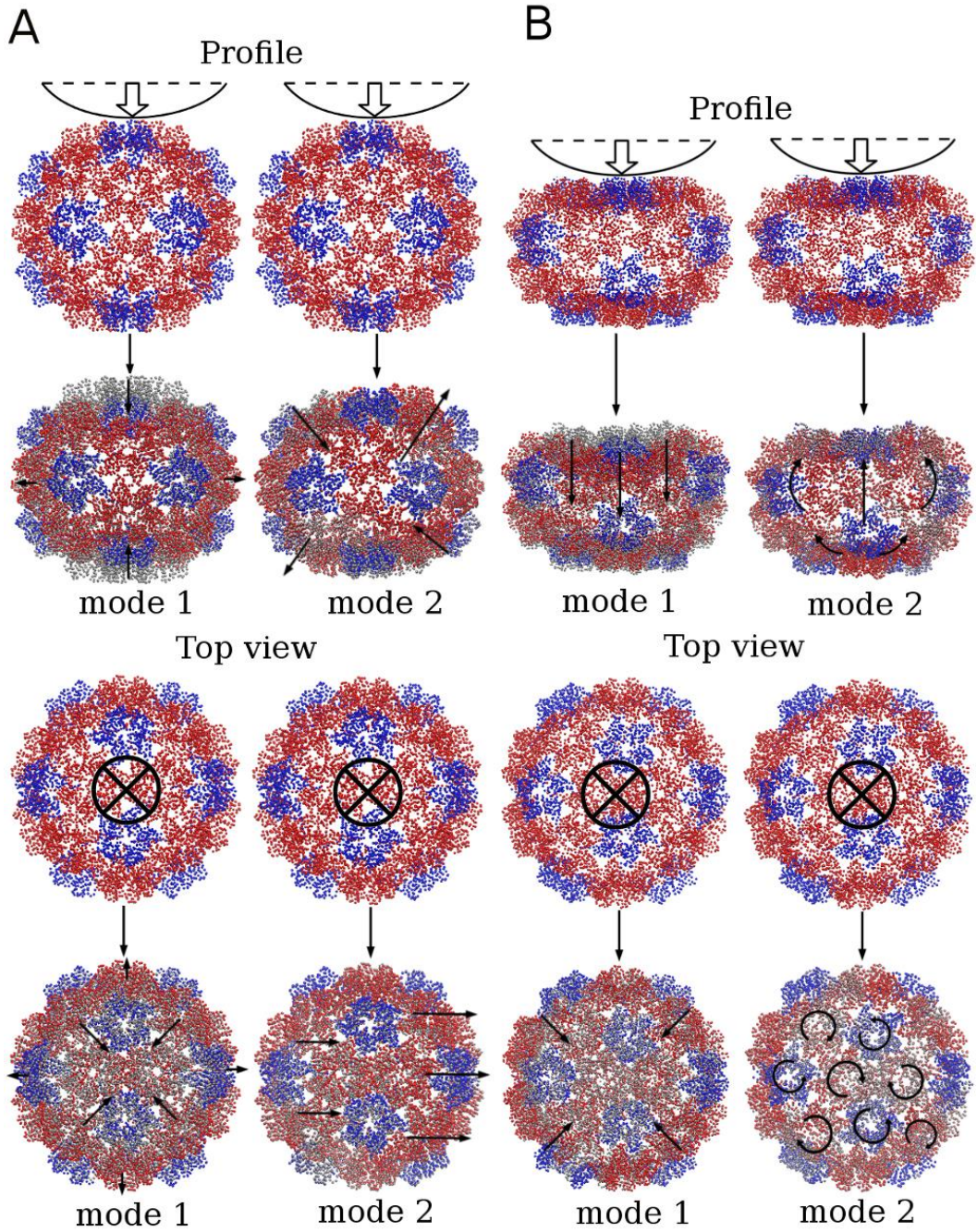

**Figure 5 (Kononova, Snijder, Brasch, Cornelissen, Dima, Marx, Wuite, Roos, Barsegov)**

# Structural Transitions and Energy Landscape for Cowpea Chlorotic Mottle Virus Capsid mechanics from nanomanipulation *in vitro* and *in silico*[*]

Olga Kononova,[†‡] Joost Snijder,[§] Melanie Brasch,[¶] Jeroen Cornelissen,[¶] Ruxandra I. Dima, [‖] Kenneth A. Marx, [†] Gijs J. L. Wuite,[§] Wouter H. Roos[§*] and Valeri Barsegov[†‡*]

[†]Department of Chemistry, University of Massachusetts, Lowell, MA 01854; [‡]Moscow Institute of Physics and Technology, Moscow Region, Russia 141700; [§]Natuur- en Sterrenkunde and LaserLab, Vrije Universiteit, 1081 HV Amsterdam, The Netherlands; [¶]Biomoleculaire Nanotechnology, Universiteit Twente, 7500 AE Enschede, The Netherlands; [‖]Department of Chemistry, University of Cincinnati, Cincinnati, OH 45221.

## Supporting Material

[*]Corresponding authors:
e-mail: Valeri_Barsegov@uml.edu, tel: 978-934-3661, fax: 978-94-3013
e-mail: wroos@few.vu.nl, tel: +31 20 59 83974, fax: +31 20 59 87991

*Running Title*: Mechanical properties of CCMV capsid

**Self-Organized Polymer (SOP) Model Parametrization:** Numerical values of $\varepsilon_n$ were determined from equilibrium all-atom Molecular Dynamics (MD) simulations in implicit water for the CCMV shell at $T$=300 K (CHARMM19 force field) (1). We employed the Solvent Accessible Surface Area (SASA) model of implicit solvation (2). We used the CCMV structure from PDB (PDB code: 1CWP) to carry out a 5 ns equilibrium simulation run for the entire capsid. First, we calculated the number of native contacts, which stabilize the native state of the capsid, based on a standard cut-off distance $R_C$=8 Å between the $C_\alpha$-atoms. The native contacts were divided into the inter-chain contacts and intra-chain contacts. There were $N_{intra} \approx 20{,}554$ intra-chain contacts in the capsid proteins, and $N_{inter} \approx 3{,}405$ inter-chain contacts at their interfaces (due to the van-der-Waals and Coulomb interactions). Next, we calculated the total energy of non-covalent interactions for each contact group. We found $E_{intra} = 25{,}898$ kcal/mol for the intra-chain contacts and $E_{inter} = 3{,}746$ kcal/mol for the inter-chain contacts. Finally, we divided the total energy by the number of contacts for each group of contacts. We obtained $\varepsilon_{intra} = E_{intra}/N_{intra} = 1.26$ kcal/mol for the intra-chain energy, and $\varepsilon_{inter} = E_{inter}/N_{inter} = 1.1$ kcal/mol for the inter-chain energy. These parameters of the SOP force field were used in the simulations of mechanical indentation of CCMV.

**Langevin simulations of CCMV indentation:** The indentation dynamics were obtained by integrating the Langevin equations for each particle position $r_i$ in the over-damped limit, $\eta \mathrm{d}r_i/\mathrm{d}t = -\partial U/\partial r_i + g_i(t)$. Here, $U$ is the total potential energy, which accounts for the contribution from the capsid conformation $U_{SOP}$ and for the interaction of the $i$-th particle with the spherical tip $U_{tip}(r_i)$ (see Materials and Methods in the main text). Also, $g_i(t)$ is the Gaussian distributed random force and $\eta$ is the friction coefficient. The Langevin equations were propagated with the time step $\Delta t = 0.08\tau_H = 20$ ps, where $\tau_H = \zeta \varepsilon_h \tau_L/k_B T$. Here, $\tau_L = (ma^2/\varepsilon_h)^{1/2} = 3$ ps, $\zeta = 50.0$ is the dimensionless friction constant for a residue in water ($\eta = \zeta m/ \tau_L$ ), and m $\approx 3 \times 10^{-22}$ g is the residue mass (3,4). Simulations of mechanical indentation were carried out at room temperature using the bulk water viscosity, which corresponds to the friction coefficient $\eta = 7.0 \times 10^5$ pN ps/nm.

**Umbrella Sampling Method:** We calculated the potential of mean force $U$ as a function of indentation depth $X$ for the process of mechanical compression of CCMV particle. In this approach, the potential energy of the system depends on $X$ (reaction coordinate), which gradually changes during the simulation, and one accumulates $\langle \partial U / \partial X \rangle$ at several values of $X$ (<...> denotes the ensemble averaging). The potential of mean force can then be estimated as $\int \langle \partial U / \partial \lambda \rangle d\lambda$ (5,6). Because the indentation depth $X$ is the distance travelled by the cantilever tip, $\partial U / \partial x$ corresponds to the indentation force ($F$), i.e. the force exerted on the tip by the virus particle. Since $F = \kappa dX$, where $\kappa$ is the cantilever spring constant and $dX$ is the tip displacement, the measurements of $F$ can be used to estimate $\partial U / \partial X$ and to calculate $U$. We performed one simulation run to collect 280 data points. In each step, the cantilever (base) was moved by 0.1 nm; next, the system was equilibrated for 0.2 ms, and then the values of the tip position were sampled for 0.2 ms to calculate $F = \partial U / \partial X$ . We used the cantilever spring constant $\kappa$= 1.4 N/m.

**Normal Mode Analysis:** In the NMA, the potential energy is expanded in a Taylor series in terms of the mass-weighted coordinates $q_i = \sqrt{m_i} \Delta x_i$, where $\Delta x_i$ is the displacement of the *i*-th particle from the energy minimum and $m_i$ is its mass. The Hessian matrix of second derivatives $H$ with the matrix elements $H_{ij} = \partial^2 V / \partial q_i \partial q_j$ (*i,j=1,2,...,3N*) carries information about equilibrium dynamics of the system in question. The NMA involves the following three steps: 1) minimization of the potential energy, 2) calculation of the Hessian matrix, and 3) diagonalization of the Hessian matrix (7). The energy minimization was performed using the steepest descent algorithm. Each residue was represented by its center of mass. The numerical calculation of the Hessian matrix was carried out using atomic forces $f_i = -\partial V / \partial q_i$, i.e. $H_{ij} = -[f_i(\bar{q} + h\bar{e}_j) - f_i(\bar{q} - h\bar{e}_j)]/2$, where $\bar{q} = \{q_i\}$, $\bar{e}_j$ is the unit vector in the direction of atom *j*, and *h* is the displacement along this direction. We used the transformation $H_{IJ} = (1/\sqrt{M_I M_J}) \sum H_{ij}$ to obtain the Hessian matrix for the centers of mass $H_{IJ}$ from the Hessian matrix for atomic displacements $H_{ij}$. Here, $M_I$ and $M_J$ are the reduced mass of residue *I* and *J*, where *I, J=1,2,...,N* and *N* is the total number of residues. To build the Hessian matrix for the displacements of the centers of mass of amino acids, and to solve numerically the eigenvalue problem **HR=ΛR**, we used an algorithm implemented in the GROMACS package (8). We determined the *I*-th eigenvalue, $\lambda_I$, and the *I*-th eigenvector $R_I$. Each eigenvector specifies a normal mode coordinate $Q_I$, i.e. $Q_I = \sum R_{IJ} q_j$, which oscillates with the characteristic frequency $\omega_I = \sqrt{\lambda_I}$. The spectrum of eigenvalues $\{\lambda_I\}$ was used to construct the spectrum of eigen-frequencies in the center-of-mass representation $\omega_I = (1/2\pi c)\sqrt{\lambda_I}$ (*c* is the speed of light).

**Essential Dynamics:** In the Essential Dynamics approach, one assumes that the most important displacements reside in a subspace of a few degrees of freedom, whereas the remaining degrees of freedom represent less important fluctuations (9). Dynamic correlations between particle positions at time *t*, $X(t)=\{X_1(t), X_2(t),..., X_N(t)\}$, and the position in the reference (equilibrium average) structure $X_0=\{X_1(0), X_2(0),..., X_N(0)\}$ can be expressed through the covariance matrix $C(t) = <(X(t) - X_0)(X(t) - X_0)^T>$, where $<...>$ denotes the ensemble averaging and the superscript *T* represents the transposed matrix. By construction, *C* is a symmetric matrix, which can be diagonalized by an orthogonal transformation *T*, $C = TLT^T$, where *L* is the diagonal matrix of eigenvalues and *T* is the matrix of eigenvectors of *C*. In the three-dimensional space, there are *3N-6* eigenvectors with non-zero eigenvalues for a system of *N* particles (excluding translations and rotations). The eigenvalue $l_I$ in the center-of-mass representation (*I =1,2,...,N*) is the amplitude of displacement $X(t) - X_0$ along the *I*-th eigenvector $t_I$. The principal coordinates $P_I(t)$ are obtained by projecting $X(t) - X_0$ onto each eigenvector $P_I(t) = t_I(X(t) - X_0)$. In the Cartesian basis, these projections are given by $X(t) = P_I(t) t_I + X_0$ (4,6).

.

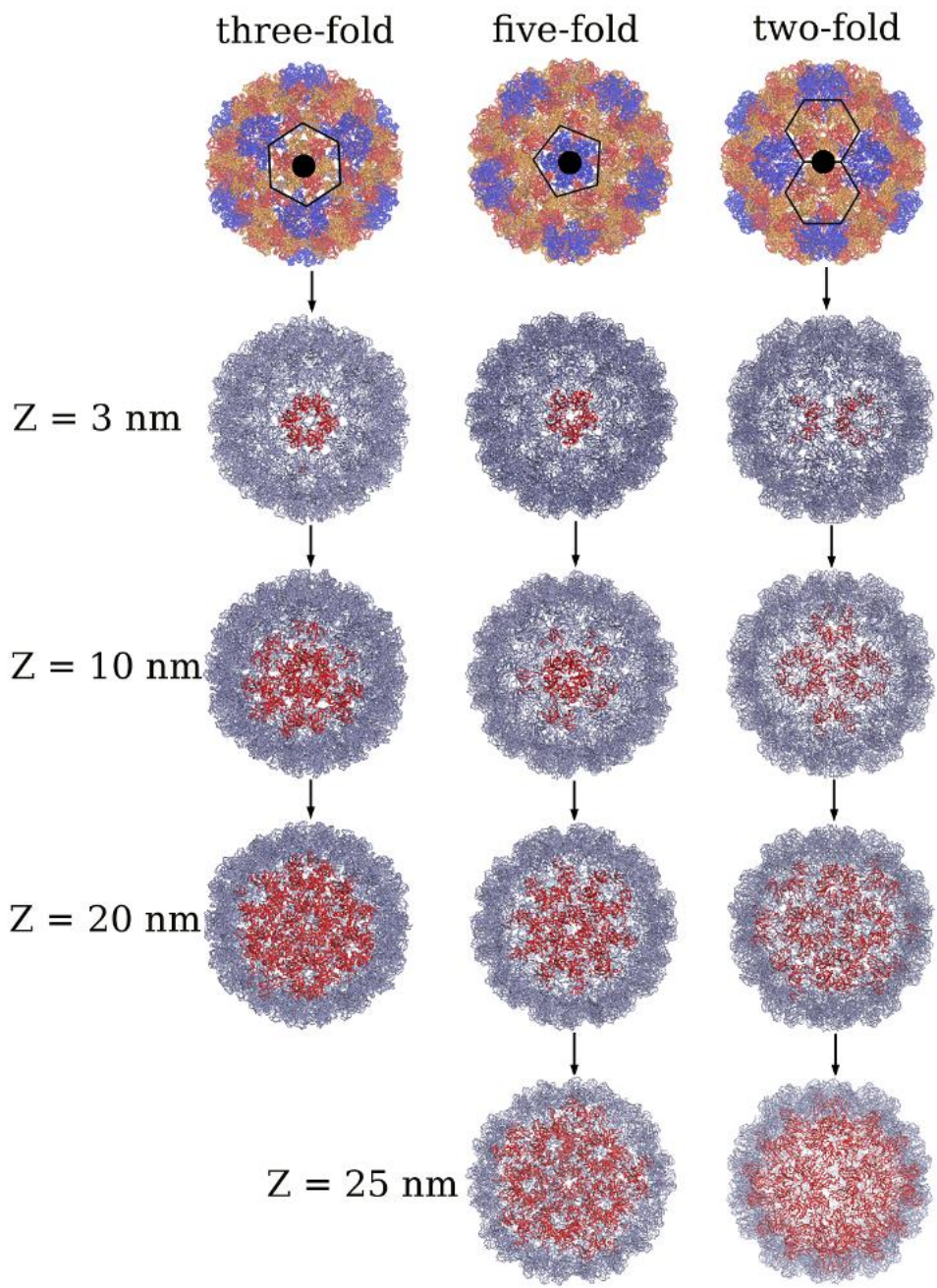

**Figure S1**. The CCMV shell: Forced indentation of CCMV along the two-fold, three-fold, and five-fold symmetry axis. The top structures show the top view of symmetry (filled black circle). Pentamers and hexamers are shown, respectively, in blue and red color. The grey structures show the tip-capsid contact area (in red) for different values of $Z = 3$ nm, 10 nm, 20 nm, and 25 nm and for different symmetry.

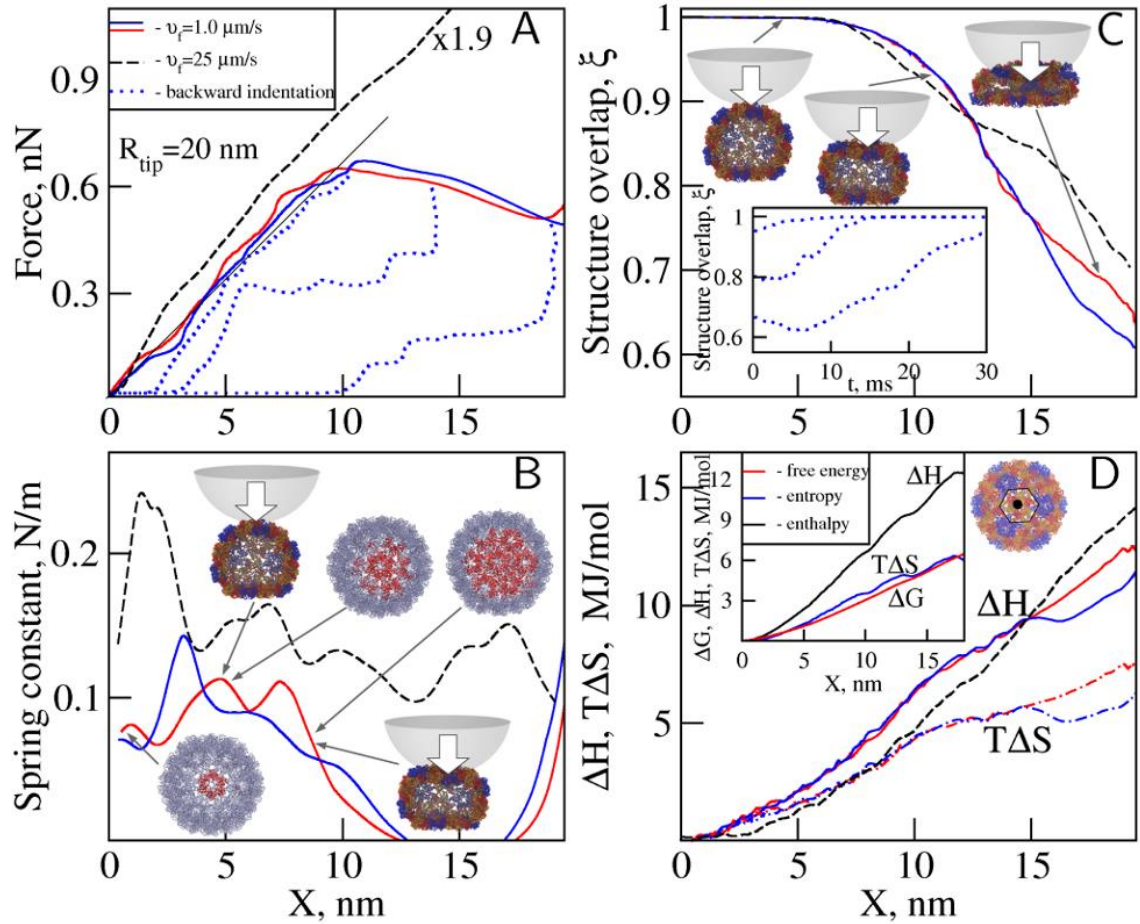

**Figure S2.** Forced indentation *in silico* of the CCMV capsid along the three-fold symmetry axis (see Fig. S1). Shown are the same quantities as in Fig. 4 in the main text, obtained using $R_{tip}$= 20 nm, $v_f$= 1.0 µm/s (red and blue curves) and $v_f$= 25 µm/s (dashed black curves), but for a different symmetry type (the top view of symmetry is shown in panel D). These results should be compared with the results of indentation along the two-fold symmetry axis (Fig. 4 in main text), and five-fold symmetry axis (Fig. S3). Also shown are the snapshots of CCMV shell (top view and profiles). The growing with indentation depth tip-capsid contact area is shown in red color (see also Fig. S1).

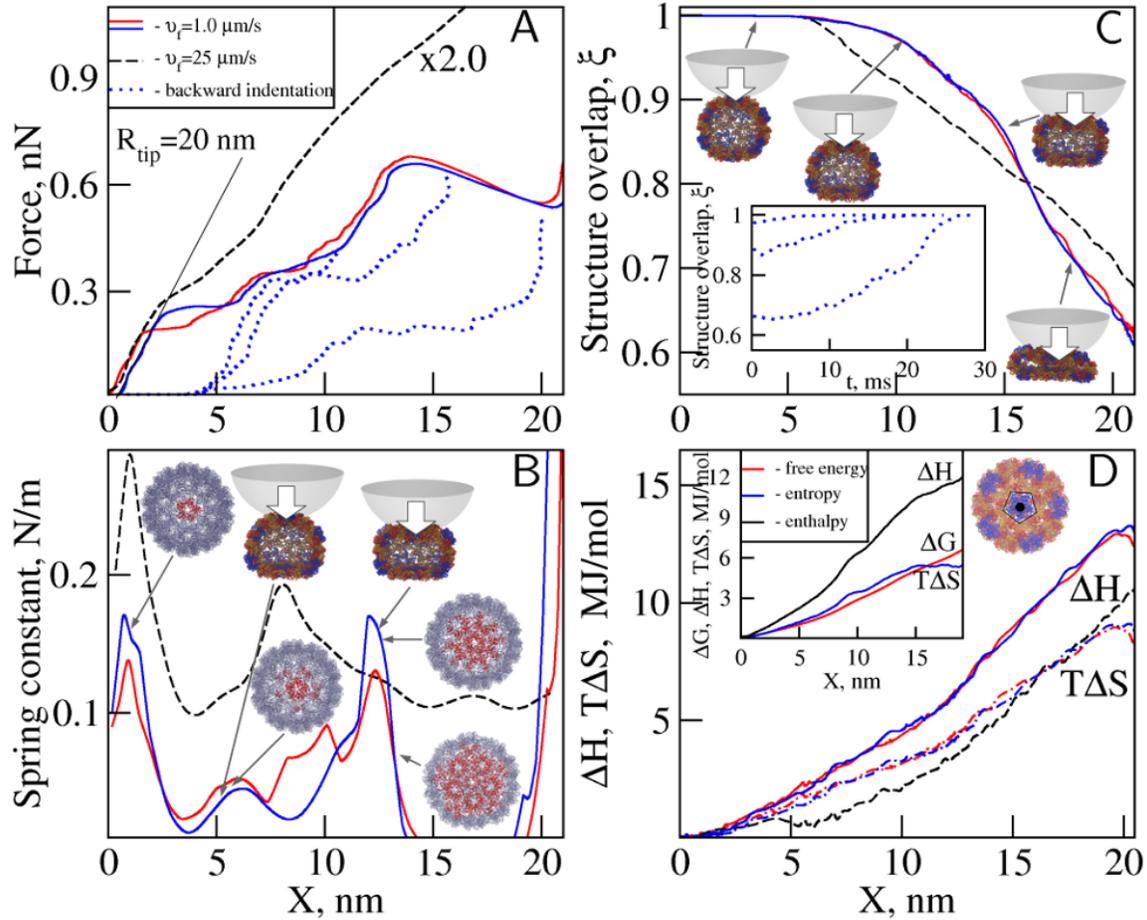

**Figure S3.** Forced indentation *in silico* of the CCMV capsid along the five-fold symmetry axis (see Fig. S1). Shown are the same quantities as in Fig. 4 in the main text, obtained using $R_{tip}$= 20 nm, $v_f$= 1.0 μm/s (red and blue curves) and $v_f$= 25 μm/s (dashed black curves), but for a different symmetry type (the top view of symmetry is shown in panel D). These results should be compared with the results of indentation along the two-fold symmetry axis (Fig. 4 in main text), and three-fold symmetry axis (Fig. S2). Also shown are the snapshots of CCMV shell (top view and profiles). The growing with indentation depth tip-capsid contact area is shown in red color (see also Fig. S1).

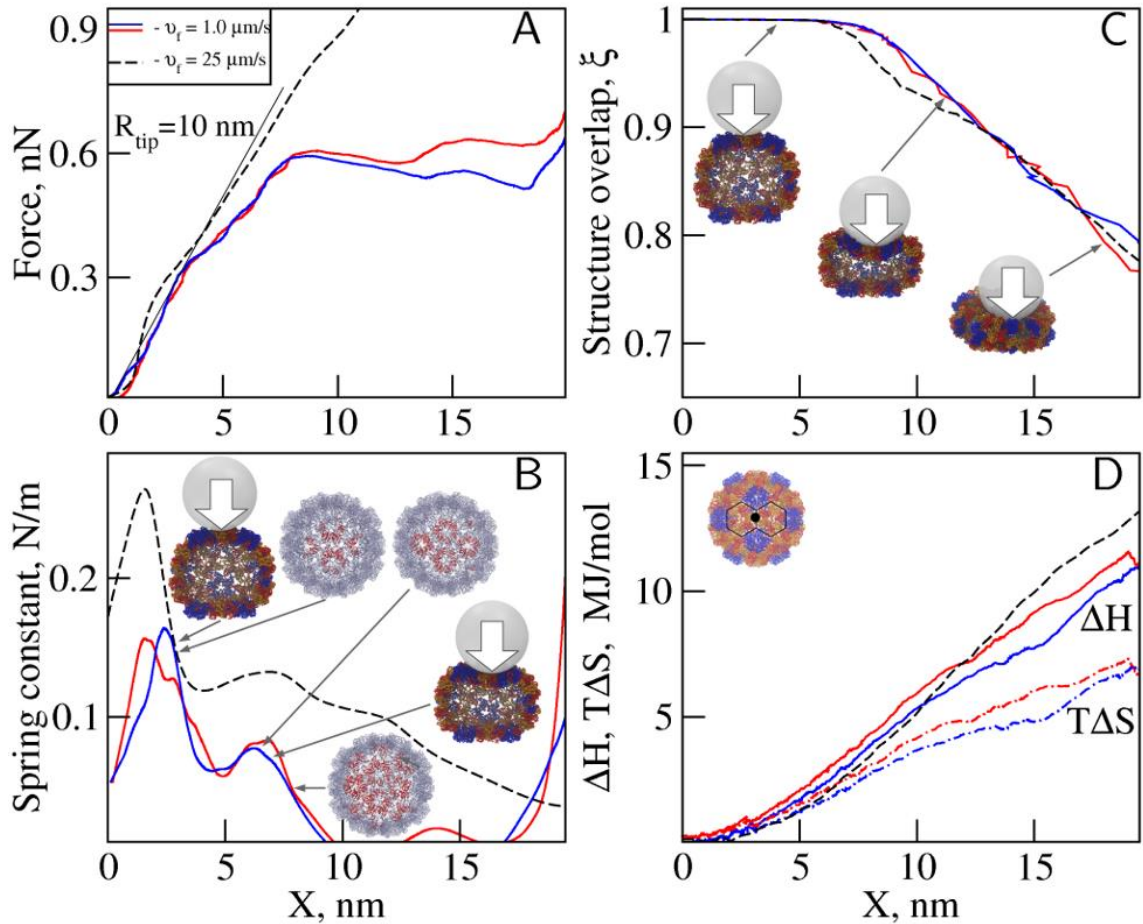

**Figure S4**. Forced indentation *in silico* of the CCMV capsid along the two-fold symmetry axis type (shown in panel D). Shown are the same quantities as in Fig. 4 in the main text but for a smaller tip of radius $R_{tip}$= 10 nm. The force peaks in the FX curves (panel **A**) are lower (~0.6 nN) compared to the results for $R_{tip}$= 20 nm, and the spring constant $k_{cap}$ now varies between 0.03 and 0.11 N/m (panel **B**). The $X$-dependence of $k_{cap}$ is still bimodal, as for $R_{tip}$= 20 nm, but the second peak at $F \approx 15$ nm is now weaker. The structure overlap $\xi$ shows that the CCMV particle in the collapsed state remains ~70-75% similar to the native (un-indented) state (panel **C**). The associated enthalpy change $\Delta H_{ind} \approx 10$ MJ/mol is slightly lower, and the entropy change $\Delta S_{ind} \approx 6$ MJ/mol is roughly the same as for the case of $R_{tip}$= 20 nm (panel **D**). Also shown are the snapshots of CCMV shell (top view and profiles). The growing with indentation depth tip-capsid contact area is shown in red color (see also Fig. S1).

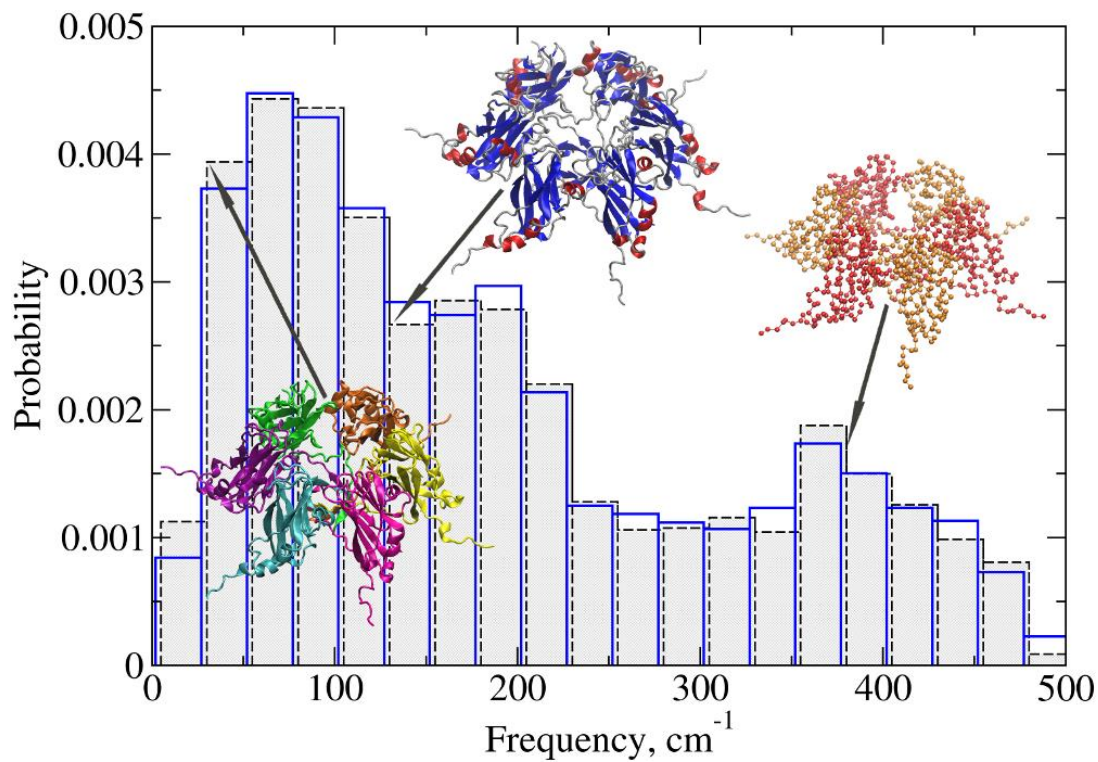

**Figure S5.** Equilibrium modes of motion of CCMV particle. The spectrum of normal eigen-frequences (histogram) for isolated single hexamers (shown by gray bars) is compared with the spectrum of eigen-frequences for the entire CCMV shell (blue bars). Structures display the scale of normal displacements from the global modes, which include displacements of chains and entire capsomers (lefthand structure) to local modes of motion of secondary structure elements (α-helices and β-strands, middle structure), and to more localized modes (vibrations of amino acid residues, righthand structure).

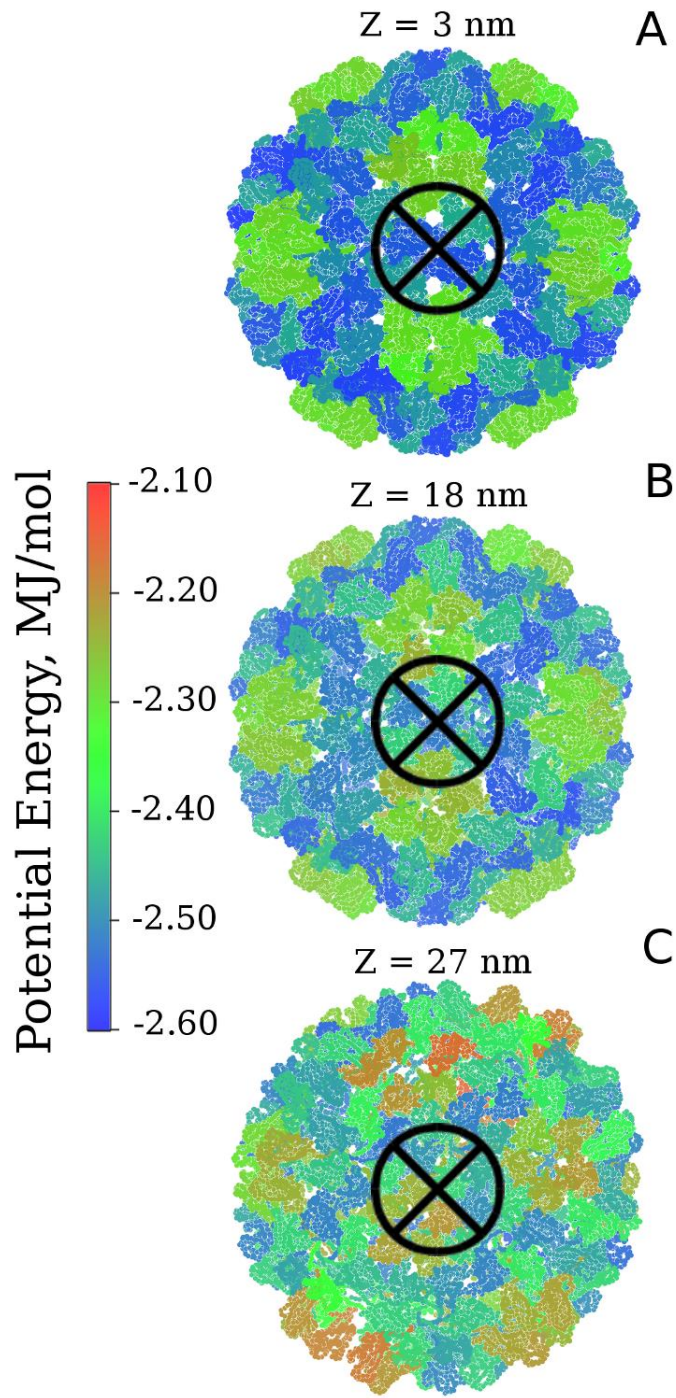

**Figure S6.** Surface map of the potential energy (color scale for $U_{SOP}$ is in the graph) for three representative structures of the CCMV shell (top view) observed in the course of deformation *in silico* at $Z = 3$ nm (panel **A**), 18 nm (panel **B**), and 27 nm (panel **C**). The direction of motion of the tip is perpendicular to the CCMV surface as indicated. The map shows a gradual increase in the potential energy of proteins forming pentamers and hexamers as changes to the global structure occur.

**Movie S1.** Mechanical nanoindentation *in silico* of CCMV shell along 2-fold symmetry axis. The light blue sphere represents cantilever tip of radius $R_{tip}$ = 20 nm, moving along -$z$ direction with constant velocity $v_f$ = 1.0 μm/s against a mica-surface (not shown).

**Supporting References:**